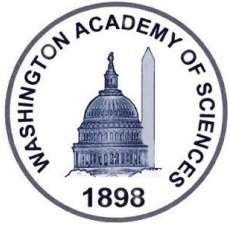

# El Bahr: A Prospective Impact Crater

Paris, Antonio, Department of Natural Sciences, St. Petersburg College, Florida
Shalabiea Osama M., Astronomy, Space Sciences and Meteorology Department, Cairo University, Egypt
Madani, Ahmed; El-Sharkawi, Mohamed, Department of Geology, Cairo University, Egypt
Davies, Evan, The Explorer Club, New York, NY

## ABSTRACT

This preliminary investigation addresses the discovery of an unidentified crater located south of the Sahara Desert between Qaret Had El Bahr and Qaret El Allafa, Egypt. The unidentified crater (hereafter tentatively named El Bahr Crater) was discovered during a terrain analysis of the Sahara Desert. El Bahr Crater, which is located at 28°40'20"N and 29°15'25"E (Southwest Al-Jiza Giza), is approximately 327 meters across, has a rim with a circumference of approximately 1,027 meters, and occupies a surface area of approximately 83,981 square meters. Preliminary spectral and topographic analysis reveal features characteristic of an impact crater produced by a hypervelocity event of extraterrestrial origin, including a bowl-shaped rim and a crater wall. No proximal and/or distal ejecta, however, are visible from Landsat imagery. Moreover, the geomorphic features, along with the fact that the El Bahr basalts are known to be rich in orthopyroxene while the surrounding basalts are not, imply an impact as the most plausible explanation. The El Bahr Crater is not indexed in the *Earth Impact Database*, and an analysis of impact structures in Africa did not identify it as either a confirmed, proposed or disproved impact crater. In collaboration with the University of Cairo, therefore, an expedition has been organized to conduct an *in-situ* investigation of El Bahr Crater, to ascertain if planar formations, shatter cones, and shock metamorphic and/or other meteoritic properties are present.

## Area of Interest

The Western Desert of Egypt (WDE) has long been targeted by geomorphologists worldwide because of its well-exposed lithology as well as a terrain and climate that predispose this region to remote sensing studies. Although many of the circular features present there have attracted scientific attention as potential impact craters, further study has determined that most of them were formed as a result of volcanic activity. The most recently verified impact crater was discovered in 2010 by a joint Egyptian-Italian team working in the extreme southern part of Egypt, east of Gebel Oweinat near Gebel Kamil. The team was able to distinguish the mineralogy of the igneous oblate structures from those formed as a result of a potential impact. The Kamil Crater (22°01'06"N, 26°05'15"E) is 45 meters in diameter, and geologically the area is Cretaceous sedimentary in origin.[1] The discovery of Kamil Crater not only positively confirmed the presence of an impact crater in the WDE but also led to further geological interest in the region, resulting in turn in the discovery of El Bahr Crater.

In the northern sector of the WDE, fault-induced folds predominate as oblate plunging anticlines. These geological structures are well exposed in the El Bahr region, east of the El Gidida iron ore mine. In contrast to the oblate structures, various circular ring structures exist, in particular El Bahr basalt. These basaltic rocks differ from the basalts exposed in the floor of the El Bahariya depression in its mineralogical composition. Ortho-pyroxene is present in El Bahr basalt but entirely absent from El Bahariya basalt.[2]

Another feature provoking interest is a prospective impact crater in the El Bahr depression (Fig. 1), located northwest of the El Bahr basalt circular structures. This crater, hereafter tentatively described as the El Bahr Crater, differs from the oblate structures that are predominant in the WDE. Analogously, the El Bahr Crater does not belong to the fault-induced folds, and the El Bahr region lies in carbonate terrain that is extruded by dark basalt features whereas the El Bahr crater does not.

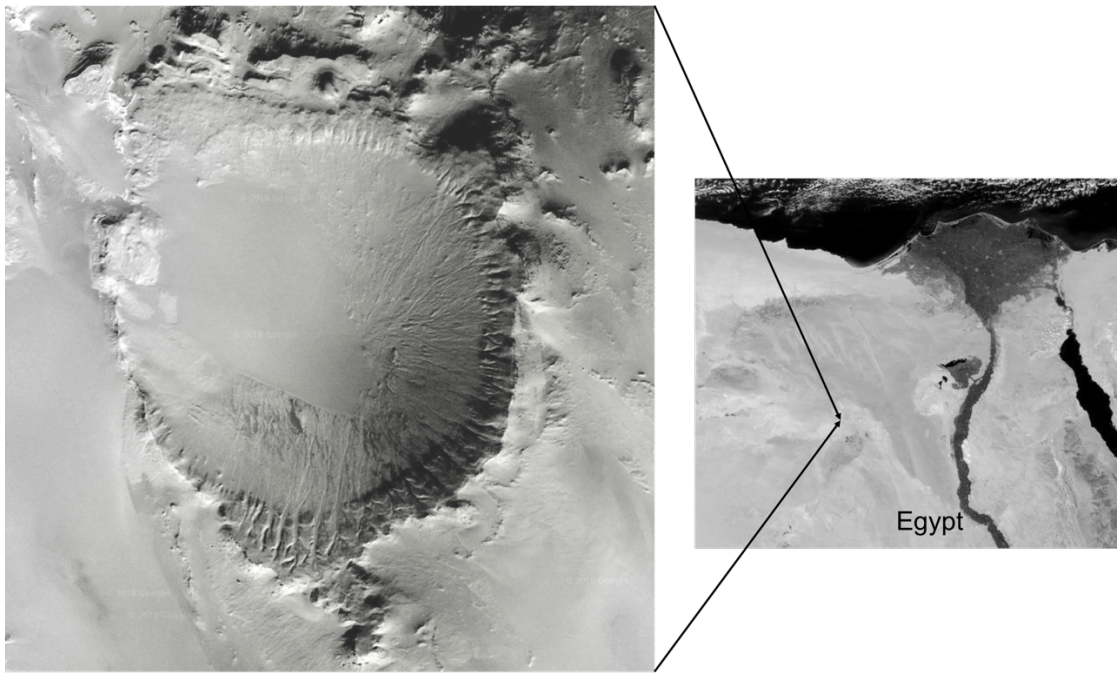

*Figure 1: Location map for El Bahr Crater, Western Desert, Egypt.*

**Preliminary Investigation of El Bahr Crater: Remote Sensing Approach**

This investigation considers all possibilities for the formation of El Bahr Crater, relying heavily on remote sensing analysis of Landsat data. Generally, the El Bahr region is dominated by carbonate rocks of the Qazzun and El Hamra Formations that were cut by basaltic volcanism. The El Bahr region was affected by Cretaceous and Miocene tectonics. The region, moreover, consists of a variety of sand dunes and sheets induced by northwest winds, and the two foremost structural elements recorded suggest that the region is composed of faults and folds. Topographically, the El Bahr Crater is located NNW along a series of strike-slip faults and folds (Fig. 2). However, it is circular-shaped and not comparably elongated like the folds in the region of interest (Fig. 2), further indicating that this crater was formed later than the fault/fold morphology in the region.

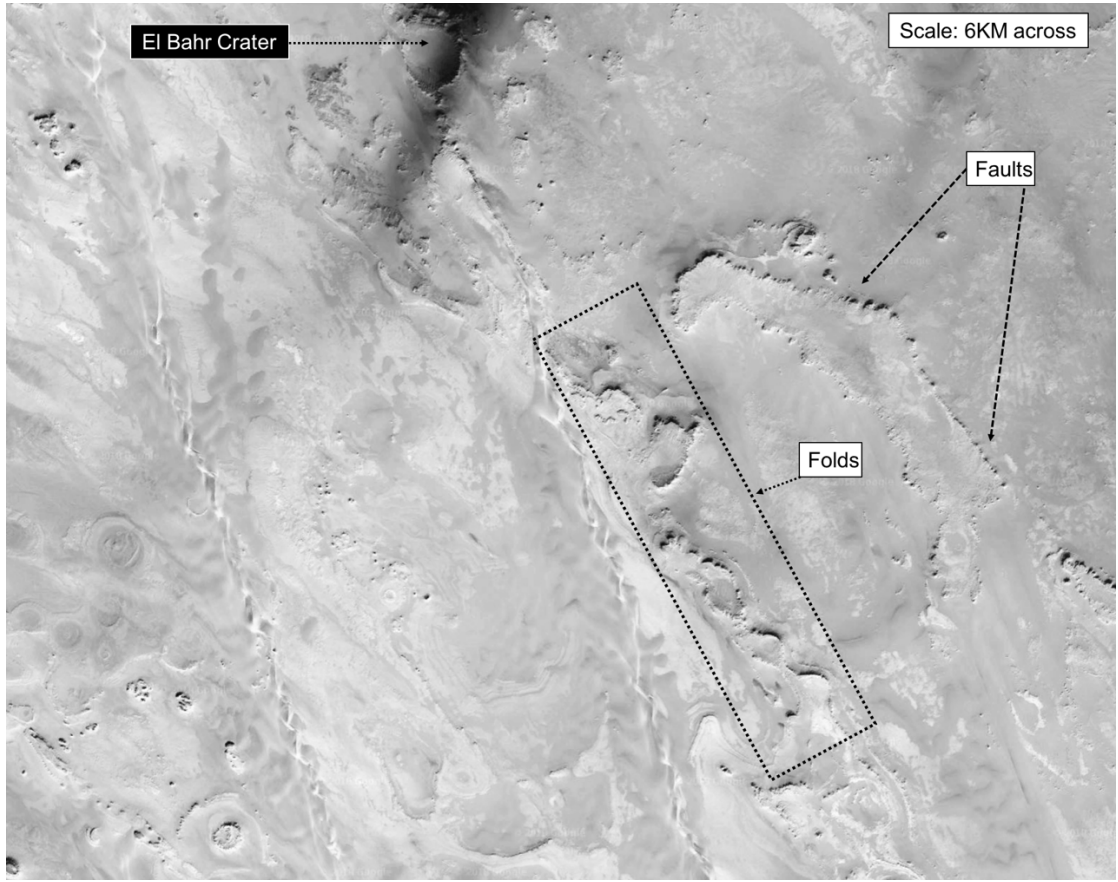

The three competing hypotheses that could explain the formation of El Bahr Crater are a man-made structural event (i.e., nuclear and/or high-power explosive test), volcanism, or a hypervelocity event extraterrestrial in origin (i.e., a meteor). The first hypothesis can be eliminated. An inspection of the morphology of El Bahr Crater from space using Landsat imagery and an examination of historical archives of the region rule out the possibility that a nuclear and/or high-power explosive test occurred in the El Bahr region. To test the remaining two hypotheses, a comparison of spectral signatures between the impact crater Kamil and El Bahr Crater was conducted. This comparison was completed through the use of data from Landsat-7's Enhanced Thematic Mapper Plus (ETM+) and digital image processing using Environment for Visualizing Images (ENVI) software.

Numerous publications have described the usefulness and sensitivity of 3/1 band ratio spectral analysis of iron mineral mapping. Accordingly, the band ratio discrimination technique was used to compare spectral signatures between the impact crater Kamil and El Bahr Crater. Figure 3 shows the laboratory spectral curves of iron oxide minerals that were downloaded from the USGS Spectral Library website.[3] Hunt and Salisbury (1970) concluded that iron minerals have low flat spectral reflectance.[4] The spectral curve of magnetite shows a very weak broad absorption feature around 1 μm, whereas spectral curves of hematite and goethite show two main absorption features around 0.6 μm and 0.9 μm.

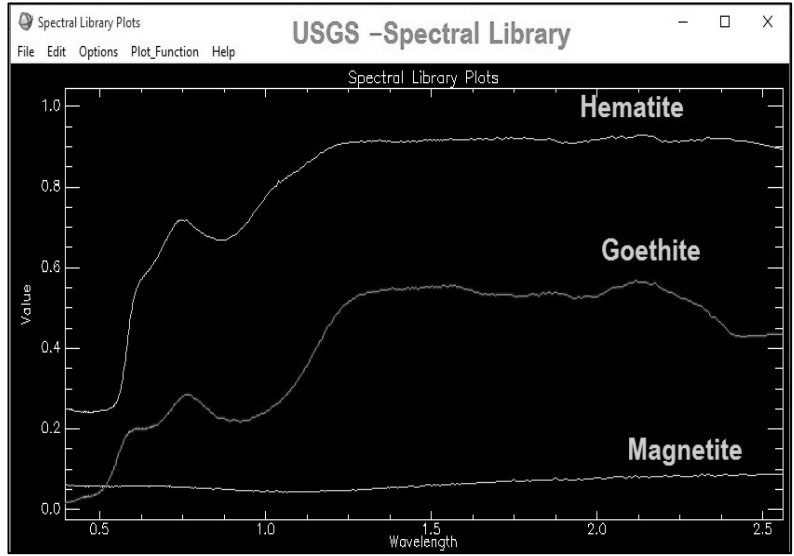

*Figure. 3: Spectral curves of hematite, goethite and magnetite minerals*

Figures 4a and 4b display 3/1 band ratio images for Kamil and El Bahr Craters, respectively. When the images are compared, Kamil Crater displays a darker signature whereas the floor and rim of El Bahr Crater exhibit bright and gray image signatures. A dark image signature, however, is recorded externally along the northern and southern rims of El Bahr Crater.

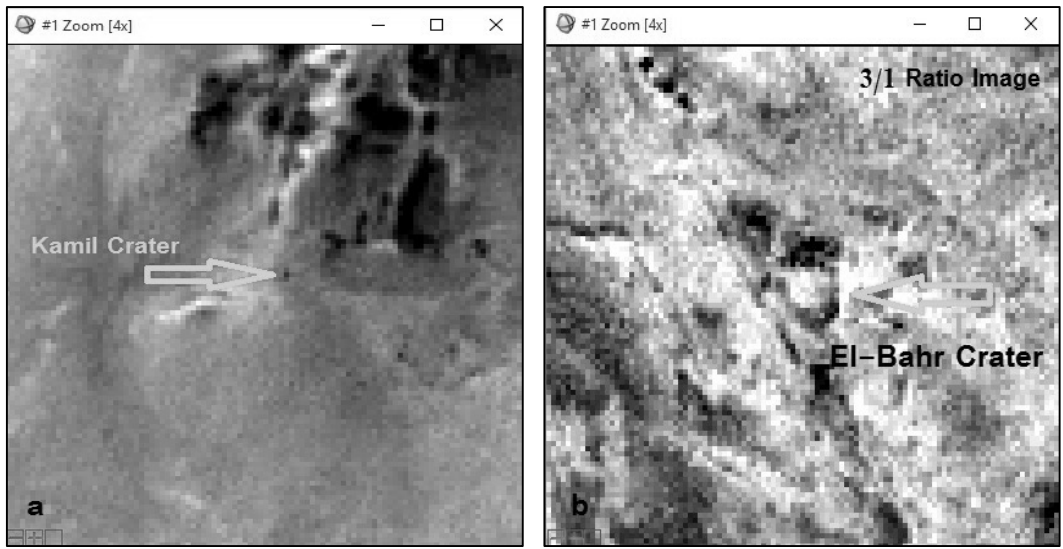

*Figure. 4: 3/1 band ratio images for: a) Kamil Crater b) El Bahr Crater.*

Figures 5 and 6 display the spectral values of 3/1 band ratio images for Kamil and El Bahr Craters, respectively. An inspection of these values revealed that (1) Kamil Crater has low value compared with the immediate surroundings and (2) the floor and rim of El Bahr Crater have high and moderate values. The only low values at El Bahr Crater are recorded outside the crater, in the direction of due north.

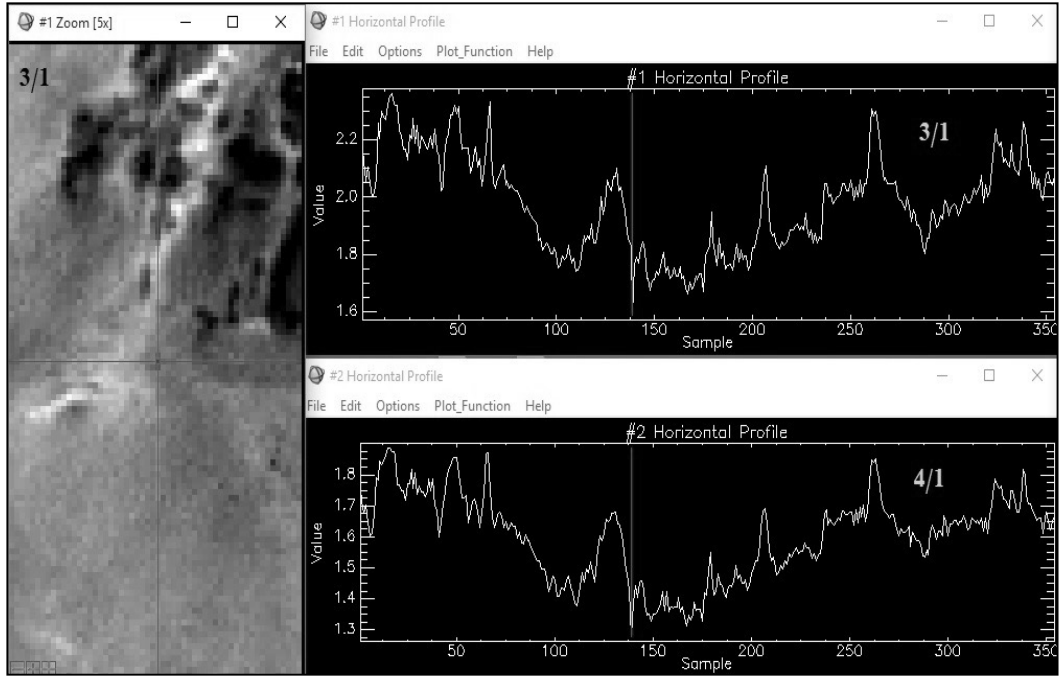

*Figure 5: 3/1 ratio image spectral values. Note that the Kamil Crater has low values in 3/1 band ratio images compared with its surroundings.*

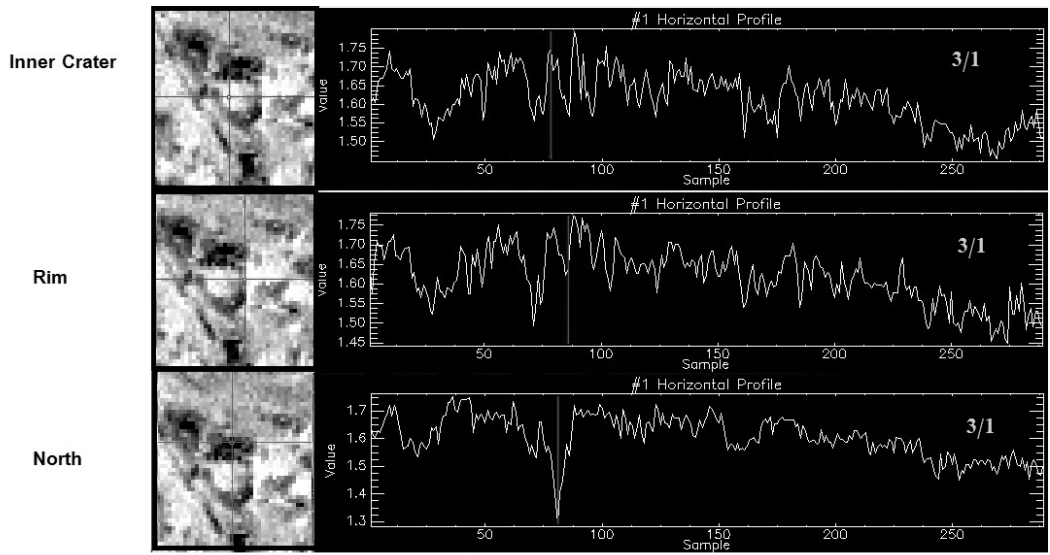

*Figure 6: 3/1 ratio image spectral values for the floor, rim, and outside regions. Note that the floor and rim of El Bahr Crater have high and moderate values compared with its surroundings. The only low value is recorded outside the crater.*

Finally, figure 7 displays two small areas near the El Bahr Crater, which have the same signature as El Bahr basalt. One is Area A in the figure and lies along the northern rim of the El Bahr Crater. The second, Area B, lies southwest of the crater and represents El Bahr basalt.

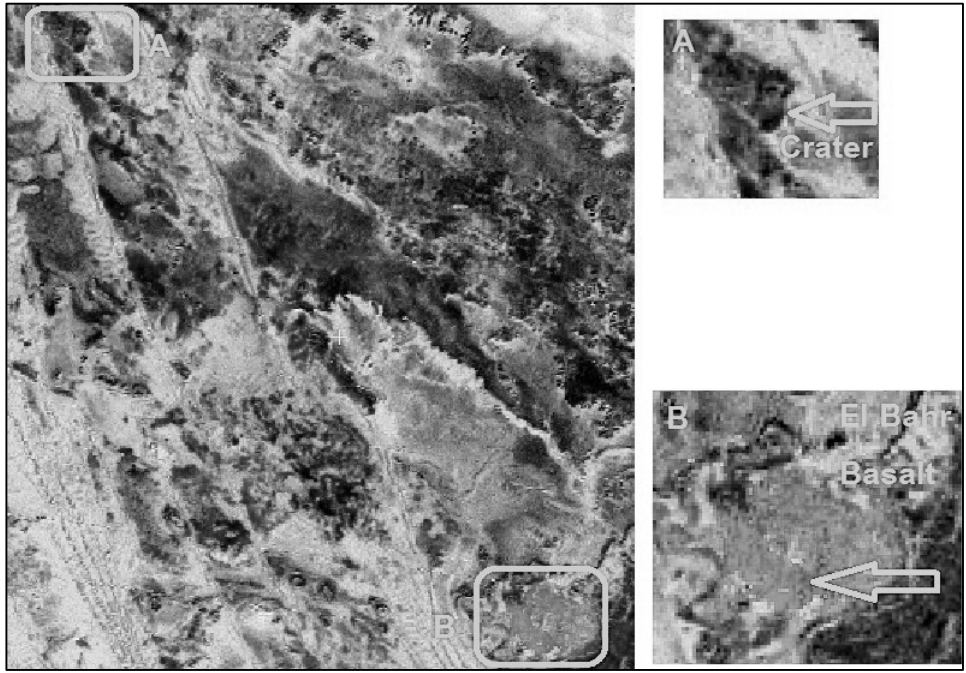

*Figure 7: False color composite band ratios image of El Bahr basalt and the proposed crater.*

## Preliminary Conclusions

Both the El Bahr basalt and the proposed crater are aligned along N30W in the El Bahr depression (Fig. 2). South of the proposed crater, all the El Bahr basalt occurs in circular features that are structurally elongated, whereas El Bahr Crater is not. The fault-induced folds in the region are doubly plunging structures that predate these elongated structures. These elongated structures conceivably date back to the Oligocene period, whereas the doubly plunging structures perhaps pertain to the Cretaceous and subsequent Miocene period. Additionally, two small areas, one to the southeast and the other along the northern rim of the proposed crater, display the same signature as the El Bahr basalt. This could indicate that basalt is likely to occur in the proposed crater. From overhead imagery, there is no apparent ejecta blanket in the surrounding landscape, nor is any such impact indexed in either the *Earth Impact Database* or the 2014 review *Impact Structures in Africa*.[5] Consequently, a ground truth survey would be helpful in examining this previously unidentified structure. If confirmed, this structure could be well-preserved evidence of one of the more recent impact events on Earth, and much could be learned from later in-depth investigations of its mineralogy and structure.

# Biographies

Antonio Paris is the Chief Scientist at the Center for Planetary Science and a Professor of Astronomy and Earth Science at St. Petersburg College, FL. Prof. Paris is a graduate of the NASA Mars Education Program at the Mars Space Flight Center, Arizona State University and the author of *Mars: Your Personal 3D Journey to the Red Planet*.

Osama M. Shalabiea is a Professor of Astrophysics at Cairo University and the Director at the Space Science Center in Cairo, Egypt. Prof. Shalabiea's research is focused on comets, nebulae and geochemistry.

Ahmed Madani is a Professor of Geology at Cairo University. Prof. Mandani has published a wide variety of publications on the geology of Western Desert, Egypt, including the spectroscopy of olivine basalts.

Mohamed El-Sharkawi is a Professor of Geology at Cairo University. His research includes geochemistry, tectonic evolution and petrography in Eastern Desert, Egypt.

Evan Davies is a geologist and a fellow of at the Royal Geographical Society and The Explorers Club. Dr. Davies is the author of *Emigrating Beyond Earth: Human Adaptation and Space Colonization* and has held a lifelong interest in comets, asteroids, and impact craters.